\documentclass[showpacs,prl,floatfix,twocolumn,amsmath,a4]{revtex4}

\usepackage{graphicx}

\begin{document} 

\title{Evidence for multiferroicity in TTF-CA organic molecular crystals}

\author{Gianluca Giovannetti$^{1}$, Sanjeev Kumar$^{2,3}$, Alessandro Stroppa$^{1}$, Jeroen van den Brink$^{3,4,5}$, Silvia Picozzi$^1$}

\affiliation{
$^1$Consiglio Nazionale delle Ricerche - Istituto Nazionale per la Fisica
della Materia (CNR-INFM), CASTI Regional Laboratory, 67100 L'Aquila, Italy \\
$^2$Faculty of Science and Technology and MESA+ Research Institute,
University of Twente, Enschede, The Netherlands \\
$^3$Institute Lorentz for Theoretical Physics, Leiden University, Leiden, The
Netherlands \\
$^4$Stanford Institute for Materials and Energy Sciences, Stanford University and SLAC, Menlo Park\\
$^5$Institute for Molecules and Materials, Radboud Universiteit, Nijmegen, The Netherlands
}

\begin{abstract}
We show by means of ab-initio calculations that the organic molecular crystal TTF-CA  is multiferroic: it has an instability to develop spontaneously both ferroelectric and magnetic ordering. Ferroelectricity is driven by a Peierls transition of the TTF-CA in its ionic state. Subsequent antiferromagnetic ordering strongly enhances the opposing electronic contribution to the polarization: it is so large that it switches the direction of the total ferroelectric moment. Within an extended Hubbard model we capture the essence of the electronic interactions in TTF-CA, confirm the presence of a multiferroic groundstate and clarify how this state develops microscopically.
 \end{abstract}

\date{\today}

\pacs{71.45.Gm, 71.10.Ca, 71.10.-w, 73.21.-b}

\maketitle

{\it Introduction} 
The search for multiferroics -- single phase materials in which  magnetism and ferroelectricity coexist --  has become an important research topic in the last few years \cite{Spaldin}. A large number of new multiferroics have been discovered, often guided by the theoretical prediction on the presence of multiferroicity \cite{eerenstein,cheong,slv}. Almost all newly discovered multiferroics are transition metal compounds where spin, lattice and charge degrees of freedom are strongly entangled. Here, we explore a new direction and extend this search towards organic systems. Synthesizing organic ferroelectrics is a rapidly evolving endeavor within the field of organics, materials that are known for their lightness, flexibility and non-toxicity \cite{tokurareview}. Finding an {\it organic multiferroic} can open up a new area of materials research with potentially mechanisms for the simultaneous occurence of magnetic and ferroelectric order different from those active in standard transition metal oxide multiferroics. Here we show with a combination of density functional theory (DFT) and model Hamiltonian calculations that the organic molecular crystal TTF-CA is ferroelectric. The magnitude of the polarization strongly depends on the presence of magnetic order, thereby providing an interdependence of magnetism and ferroelectricity. To the best of our knowledge this  is the first theoretical prediction for multiferroicity in an organic material.

\begin{figure}
\centerline{
\hspace{0cm}\includegraphics[width=.65\columnwidth,angle=0]{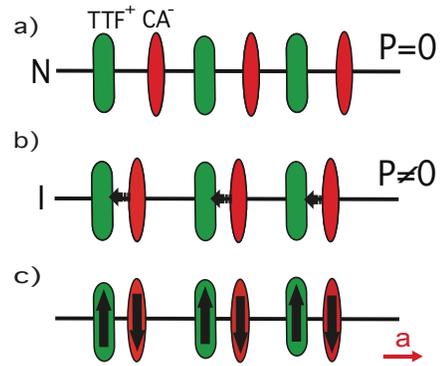}}
\caption{Schematic arrangment of TTF and CA molecules along the $a$ axis in a) the neutral (N)  b) ionic (I) structural phases with finite spontaneous polarization, 
where arrows indicate the displacement of CA towards TTF due to the dimer formation, and  c) AF magnetic structure in the ionic phase.}
\label{fig1}
\end{figure}

{\it Neutral-Ionic Transition}
TTF-CA (tetrathialfulvalene-$p$-chloranyl) is a charge transfer (CT) salt with stacks of alternating donor (TTF) and acceptor (CA) molecules. This material has been studied over the last few decades because of its prototypical neutral-ionic  transition (NIT) \cite{tokurareview, torrance}. At ambient pressure, TTF-CA undergoes a first order NIT at a critical temperature of $\sim$ 81 K\cite{MHLemmeCailleau}. The crystal symmetry is lowered from P12$_1$/\textit{n}1 (centrosymmetric) for the neutral state to P1\textit{n}1 (non-centrosymmetric) for the ionic phase \cite{LeCointe}. This symmetry breaking changes the position of TTF and CA molecules within the plane, causing the formation of pairs of long and short bonds along the stacking axis  $a$ (see Fig. \ref{fig1}). The formation of dimers is reflected in the dielectric response, which shows that the NIT in TTF-CA is associated with a first-order ferroelectric phase transition \cite{Soos,Okamoto}.  Both the non-polar and polar states of TTF-CA are electronically charged
\cite{Tokura1,Girlando}.

{\it Ab initio electronic structure}
We perform density functional theory calculations \cite{KohnHohenberg} using the generalized gradient
 approximation to exchange-correlation functional according to Perdew-Becke-Erzenhof (PBE) \cite{PBE}.
 The electronic structure is computed using the projector augmented wave method (PAW)\cite{Blochl}, 
 as implemented in the Vienna ab-initio Simulation Package (VASP)\cite{VASP,notaVASPparameters}.
The electronic contribution to the electric polarization  is calculated by the Berry phase (BP) 
  method developed by King-Smith and Vanderbilt\cite{BerryPhase}.
  
Calculations for the high temperature centrosymmetric structure (space group  P12$_1$/\textit{n}1 \cite{LeCointe}) provide a simple picture for the electronic structure: the TTF-HOMO (Highest Occupied Molecular Orbital) and CA-LUMO (Lowest Unoccupied Molecular Orbital) of the four molecules hybridize, giving rise to two bonding and antibonding  bands around the Fermi level. This situation in essence corresponds to an \textit{half-filled band system}. Electronic charge is donated from TTF to the CA, through occupation of the bonding bands \cite{Katan,Oison}.  The bandwidth of the occupied manifold is $W$=1.0 eV, the system is  \textit{metallic} and it shows a quasi one-dimensional character along the stacking axis $a$ as is highlighted in the band structure results by the dispersion along the $\Gamma$-X direction of the Brillouin zone \cite{Katan,Oison}. 

\begin{figure}
\centerline{\includegraphics[width=.99\columnwidth,angle=0]{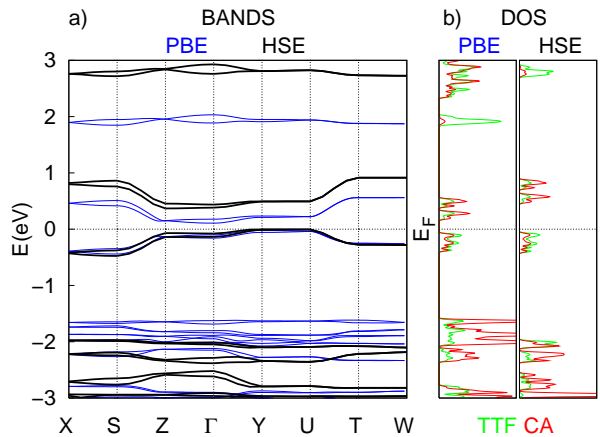}}
\caption{ Non-spin polarized band structure of P1n1 experimental structure of Ref. \cite{LeCointe}
along high simmetry directions (labelled as: X$=$($\frac{1}{2}$,0,0),
S$=$($\frac{1}{2}$,0,$\frac{1}{2}$), Z$=$(0,0,$\frac{1}{2}$), $\Gamma$=(0,0,0),
Y$=$(0,$\frac{1}{2}$,0), U$=$(0,$\frac{1}{2}$,$\frac{1}{2}$), T$=$($\frac{1}{2}$,$\frac{1}{2}$,$\frac{1}{2}$), W$=$($\frac{1}{2}$,$\frac{1}{2}$,0))
and density of states (DOS) projected onto  TTF and CA molecule and 
 calculated using either  PBE (left) or HSE (right) functional. }
\label{fig2}
\end{figure}

Using the low-symmetry structure of TTF-CA (space group P1\textit{n}1 \cite{LeCointe}) the system opens a very  small gap $\Delta$ $\sim$ 50 meV, with the formation of dimers leading to a negligible change in the bandwidth \cite{Aldane}. The corresponding PBE band structure is shown in Fig. \ref{fig2}.  Semi-local functionals fail to describe correctly some of the most chemically significant orbitals of organic molecules, in particular the  HOMO and LUMO molecular orbitals. This defect can partly be traced back to a self-interaction error, \textit{i.e.} the spurious Coulomb interaction of an electron with itself \cite{Kronik}. Hybrid functionals, which mix a fraction of the Fock exchange with the exchange density functional have shown to largely overcome these shortcomings for molecules\cite{Scuseria_review} and molecular crystals \cite{Perger}. We thus expect with hybrid functionals a better physically description of molecular crystals, also in the light of their cumulative "track record" in organic chemistry \cite{Kronik}. We note that the DFT+U method \cite{LDAU}, often used to deal with strong on-site Coulomb correlations in transition metal oxides, is not directly applicable in molecular crystals. The reason is that in molecules the localized orbitals are multi-center rather than single-center, since the basis set correspond to orthonormal molecular orbitals instead of orthonormal atomic orbitals. Here we used the Heyd-Scuseria-Ernzerhof hybrid functional (HSE) \cite{hse} which does impressively well for extended solid state systems \cite{Scuseria_review,Kressehybrids,Stroppahybrids}. 

As opposed to PBE results, the Non-Magnetic (NM) HSE ground state of the ionic state P1\textit{n}1 already has a gap: $\Delta$=0.4 eV\cite{notahse}. The corresponding band structure obtained with HSE is shown in Fig. \ref{fig2}. We also performed spin polarized HSE calculations and find that the sublattice of TTF and CA molecules stacked along the $a$ axis become antiferromagnetically (AF) ordered, with a magnetization of ~0.4 $\mu_{B}$ per molecule (see  Fig. \ref{fig1} c))\cite{noteAF}.  The energy gain of the magnetic state with respect to the non-spin polarized state is 80 meV per unit cell \cite{notaAFDFT}. The electronic gap in the magnetic groundstate is $\Delta$=0.5 eV. 

\begin{figure}
\centerline{
\includegraphics[width=.9\columnwidth]{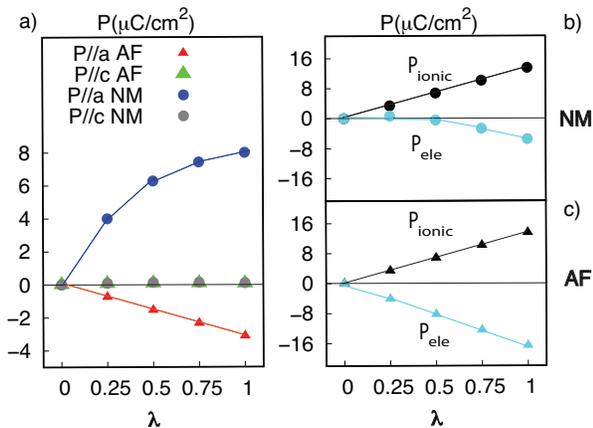}}
\caption{(a) HSE Ferroelectric polarization along $a$ and $c$ axis ($P_a$ and $P_c$) for AF and NM states
 along the path connecting the P12$_1$/n1 ($\lambda$=0) and P1n1 ($\lambda$=1) crystal structure. (b) and (c) Decomposition of $P$ into electronic  ($P_{ele}$) and ionic ($P_{ionic}$) contributions for the NM (panel b) and AF (panel c) states.}
\label{fig3}
\end{figure}

We evaluate the ferroelectric polarization $P$ in both the NM and AF states. We characterize the adiabatic path of ionic displacements from the paraelectric (P12$_1$/\textit{n}1) to the ferroelectric (P1\textit{n}1) structure by $\lambda$. The parameter $\lambda$ is therefore a measure for the amount of dimerization of TTF and CA molecules (see Fig. \ref{fig1}). In the NM state, we find a net polarization of $P=8.0\mu C/cm^2$, which reduces to $P=3.5 \mu C/cm^2$ in the AF state (see Fig. \ref{fig3}a). The polarization is in the $a-c$ plane \cite{Onoda,notasimmetry} (see Fig. \ref{fig3}a)), \textit{i.e.} ${\bf P}=P_{a}\hat{\bf a}+P_{c}\hat{\bf c}$, with an almost negligible $P_c$ component. It is remarkable that $P_{a}$ depends strongly on both the spin state and $\lambda$: $P^{NM}_{a}$ is much larger than P$^{AF}_{a}$. The decomposition of the total polarization into an ionic and electronic part, $P=P^{ele}+P^{ionic}$, elucidates this observation, see Fig.~\ref{fig3}b). These two components both increase in magnitude along the adiabatic path ({\em i.e.} as a function of $\lambda$) but are always opposite to each other.  Such a partial cancellation of electronic and ionic polarizations is also encountered in some inorganic materials, for example in the proper ferroelectric BaTiO$_3$, and in improper multiferroic materials such as HoMn$_2$O$_5$\cite{Giovannetti08}, where the polarization is magnetically driven.

Compared to the NM situation, in the AF state 
$P^{ele}_{a}$ becomes much larger due to superexchange driving the formation of a larger dipole moment in the electron cloud. As $P^{ele}_{a}$ is opposite to the $P^{ionic}_{a}$ (not affected by the magnetic ordering), an increase in the AF state of the former first causes a reduction in the magnitude of $P$, but when it becomes larger than $P^{ionic}_{a}$, it ultimately causes the magnetic ordering to flip the direction of total $P$  (see Fig. \ref{fig3}a)) -- this is what we find to occur in TTF-CA.  

{\it Model Hamiltonian}
In order to understand the mechanism leading to structural and ferroelectric transitions, we model TTF-CA by an extended Hubbard model on a one-dimensional lattice with the Hamiltonian \cite{nagaosa,del-Freo, furusaki, painelli}:
\begin{eqnarray}
&& H = -t \sum_{i \sigma}
[1 + \alpha (u_i - u_{i+1})] \left ( c^{\dagger}_{i \sigma} c^{~}_{i+1 \sigma} + H.c. \right ) \nonumber \\
&& + \delta \sum_{i} (-1)^i n_{i}
+ U \sum_{i} n_{i \uparrow} n_{i \downarrow} + 1/2 \sum_i u_i^2.
\end{eqnarray}
Here, $c^{}_{i \sigma}$ and $c^{\dagger}_{i \sigma}$ are annihilation and creation operators for electrons at site $i$ with spin $\sigma$, $t$ denotes the bare hopping strength and $u_i$ are the distortions along the chain of the TTF and CA ions from their equilibrium position.  The electron-phonon coupling parameter is $\alpha$, which favors Peierls dimerization and $\delta$ is the magnitude of the ionic potential arising due to inequivalency of the TTF and CA and $U$ denotes the Hubbard repulsion strength. 

The hopping parameter is $t=0.2 eV$ as extracted from a fit of the PBE band structure to one dimensional tight-binding chain and the on-site difference is  $\delta$=0.06 eV \cite{Katan,Oison}.  The Hubbard $U$ is the screened energy cost required to double occupy a molecular site and can be evaluated from total energy of molecular ions TTF and CA and the polarization energy 
\cite{giovannetti,tosatti,capone,Brocks04,Koch,Brink95,Meinders95}. We calculated the variation of the eigenvalues of HOMO of TTF and LUMO of CA and their polarization energy of the charged molecules placed inside a cavity of an homogeneous dielectric medium with dielectric constant of 3 using the SS(V)PE model \cite{Chipman00}. We estimate U$_{TTF}$=2.3 eV, which is in
good agreement with the value of Ref. \cite{Koch} and U$_{CA}$=2.7 eV \cite{giovannetti,tosatti,capone}. Note that this procedure does not account for the role of hybridization of HOMO and LUMO and dipole formation of the different molecular TTF and CA sites; it thus gives an upper bound on the value for the strength of electron-electron interactions. Moreover close to NIT the screening is enhanced by the diverging dielectric constant \cite{Soos,Okamoto}, decreasing the effective Coulomb interactions. In this respect, we remark that, by performing accurate G$_{0}$W$_{0}$ calculations\cite{kressegw} for the crystal structure of TTF-CA in the P1\textit{n}1 symmetry we obtain an upper-bound for the  quasiparticle gap of 1 eV \cite{notaGW}, pointing to a value of $U$ considerably smaller than that estimated above. All this suggests the importance of hybridization and screening in the mechanism driving the NIT.

Employing a Hartree-Fock decomposition for the Hubbard term, we solve the above model self-consistently for different initial conditions  and retain the lowest energy solutions. 
Deviation from the undistorted state,  {\em i.e.} dimerization,
is realized via a staggered pattern of distortions $u_i$. Therefore we set: $u_i = u_0 (-1)^i$ and treat $u_0$ as a variational parameter.
\begin{figure}
\centerline{
\hspace{0cm}\includegraphics[width=.98\columnwidth,clip = true]{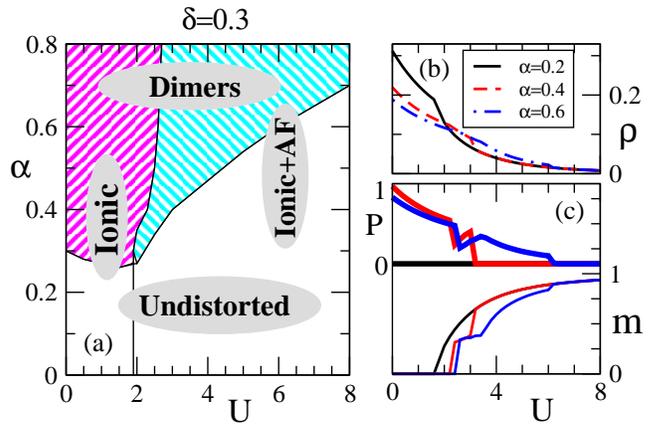}}
\caption{ (Color online) (a) Phase diagram showing undistorted to dimerized, and NM to AF transitions. The shaded region on left (right) indicates the ferroelectric (multiferroic) state. (b) Ionicity parameter $\rho = n_{TTF} - n_{CA}$ as a function of $U$ for different values of $\alpha$.
(c) The polarization computed within point charge model, and the magnetic moment as a function of $U$ for the same values of $\alpha$ as in (b).}
\label{fig4}
\end{figure}
In Fig. \ref{fig4}a, we show the groundstate phase diagram in the parameter space of $\alpha/t$ and $U/t$. On basis of the DFT calculations we fix the value of $\delta$ to $0.3 t$. The phase diagram essentially consists of two phase boundaries crossing each other: (i) non-magnetic ionic state to AFM ionic state with increasing $U$, and (ii) an undistorted state to a Peierls dimerized state with increasing $\alpha$. The ionic+dimer state is ferroelectric, shown as the
shaded region in the phase diagram. The ionicity $\rho$, here defined as the difference between the charge density at the TTF and the CA sites, arises due to the presence of the on-site energy difference $\delta$. It decreases monotonically with increasing $U$ (see Fig. \ref{fig4}b). The ionic state with dimerization of TTF-CA lacks inversion symmetry and is therefore ferroelectric. We plot the polarization calculated within a point charge model and the local magnetic moment as a function of $U$ in Fig. \ref{fig4}c. The polarization vanishes in the undistorted state due to the inversion symmetry. For $\alpha=0.4$ and $\alpha=0.6$ polarization reduces with increasing $U$ in the non-magnetic state and shows a sudden drop at the transition from a non-magnetic to AFM state due to a reduction in the dimer distortion $u_0$. The magnetic moment shows a behavior that correlates
with the polarization.  This presents an interesting case where the magnitude of the polarization is controlled by the nature of the magnetic state. We find a gap $\Delta \sim 2t$ in the DOS for $\delta/t=0.3$, $U/t=3$, and $\alpha/t=0.4$, which is fully consistent with DFT.

In conclusion, we performed first-principles calculations to understand the electronic and ferroelectric properties of the TTF-CA molecular crystal. We find that electronic correlations induce magnetism, which in turn control the ferroelectric polarization in the material. The polarization is present even in the non-magnetic state; however,  both the magnitude and direction of the polarization are very sensitive to the magnetic ordering. Using an extended 1D Hubbard model, we show that indeed, both the ferroelectric and the multiferroic states can coexist in the model and the calculation of the polarization within a point charge model qualitatively agrees with the results of first-principles calculations. Our theoretical results and predictions provide guidance for future experimental investigations on multiferroicity in the organic molecular crystal TTF-CA.

The research leading to part of these results has received funding from the European Research Council under the European Community 7$^{th}$ Framework Program (FP7/2007-2013)/ERC Grant Agreement No. 203523,
from NanoNed and FOM. The authors thank Prof. Y. Tokura and Dr. G. Kresse for a careful reading of the manuscript. Computational support by CASPUR supercomputing center (Rome) is acknowledged.


\begin{thebibliography}{99}
\bibitem{notaVASPparameters} In all the self-consistent calculations we used
a cutoff energy of 400 eV and a sampling grid of $6 \times 6 \times 4$
$k$-points with a Gaussian broadening $\sigma$=02 eV
For HSE spin polarized and G$_0$W$_0$ calculations the sampling grid was reduced to $4 \times 4 \times 2$ and $3 \times 3 \times 1$ respectively.
\bibitem{notahse} HSE calculations increase the band
  gap. The same was found for the HOMO-LUMO gap on single molecules by
  performing calculations with the GAMESS package \cite{GAMESS} with B3LYP
  functional.
\bibitem{noteAF} The exchange coupling parameter along the $c$ direction is very small, such that 
 the two possible AF configurations within the TTF-CA unit cell are degenerate in energy.
\bibitem{notaAFDFT} The AF magnetic state  could not be stabilized within bare DFT.
\bibitem{notasimmetry} In the NIT from P12$_1$/n1 to P1n1, the 2$_1$ symmetry is removed
  and the orientation of P is then consistent with this symmetry lowering.
\end{thebibliography}
\end{document}